\documentclass[prd,aps,nofootinbib,superscriptaddress]{revtex4}
\usepackage{epsfig}
\usepackage{amsmath, slashed}
\usepackage{xcolor}
\usepackage{appendix}
\usepackage{dsfont}

\usepackage[normalem]{ulem}

\usepackage{bm}
\usepackage{amsfonts}

\definecolor{pine}{rgb}{0.3, 0.5, 0.3}

\allowdisplaybreaks
\begin{document}

\title{Kinematical higher-twist corrections in $\gamma^* \gamma \to \pi \pi $ \footnote{Presented by Qin-Tao Song at DIS2022: XXIX International Workshop on Deep-Inelastic Scattering and Related Subjects, Santiago de Compostela, Spain, May 2-6 2022}}
              
\author{C\'edric Lorc\'e}
\affiliation{CPHT, CNRS, Ecole Polytechnique, Institut Polytechnique de Paris, 91128 Palaiseau, France}

\author{Bernard Pire}
\affiliation{CPHT, CNRS, Ecole Polytechnique, Institut Polytechnique de Paris,  91128 Palaiseau, France}

\author{Qin-Tao Song }
\affiliation{CPHT, CNRS, Ecole Polytechnique, Institut Polytechnique de Paris,  91128 Palaiseau, France}
\affiliation{School of Physics and Microelectronics, Zhengzhou University,  Zhengzhou, Henan 450001, China}


\date{\today}

\begin{abstract}
We apply the Braun-Manashov technique to improve the description of  $\gamma^*(q_1) \gamma(q_2) \to M(p_1) M(p_2)$ amplitudes at large $Q^2=-q_1^2$ and small $s=(q_1+q_2)^2$. We derive the kinematical higher-twist contributions of order $s/Q^2$ and $m^2/Q^2$ to the helicity amplitudes and estimate their sizes in the kinematics accessible at Belle  and Belle II.
Since pion GPDs cannot be directly measured by experiment, $\pi \pi$ GDAs are the best way to investigate the 
 energy-momentum tensor form factors for pions.

\end{abstract}

\maketitle

\section{Introduction}

Since generalized distribution amplitudes (GDAs) \cite{Muller:1994ses, Diehl:1998dk,Polyakov:1998ze}  are hadronic matrix elements of the same  bilocal quark (or gluon) operator on the light-cone as the operator entering the definition of generalized parton distributions (GPDs) \cite{Diehl:2003ny}, the techniques developed by Braun and Manashov ~\cite{Braun:2011dg,Braun:2011zr,Braun:2011th} to separate kinematical and dynamical contributions in the product of two electromagnetic currents
$T \{j_{\mu}^{\text{em}}(z_1x)j_\nu^{\text{em}} (z_2x) \}$ and applied to the deeply-virtual Compton scattering (DVCS) reaction \cite{Braun:2012bg} can be used in the description of the reaction
\begin{equation}
e(k_1) \gamma \to e' (k_2) M_1(p_1) M_2(p_2)
\label{process}
\end{equation}
accessible in $e^+ e^-$ collisions.
GDAs can be accessed in these reactions  \cite{Diehl:2000uv} in the kinematical range where $Q^2= -(k_2-k_1)^2$ is large but $s = (p_1+p_2)^2$ is much smaller than $Q^2$. They have already been the subject of careful studies at Belle \cite{Belle:2015oin, Kumano:2017lhr}. 

The kinematical corrections considered here come from two types of operators, namely 
1) the subtraction of traces in the leading-twist operators and 2) the higher-twist operators which can be reduced to the total derivatives of the leading-twist ones.
The kinematical corrections in DVCS can be considered as a generalization of the target mass corrections in deep inelastic scattering \cite{Nachtmann:1973mr}.

\section{Kinematics and generalized distribution amplitudes }

\begin{figure}[htp]
\centering
\includegraphics[width=0.6\textwidth]{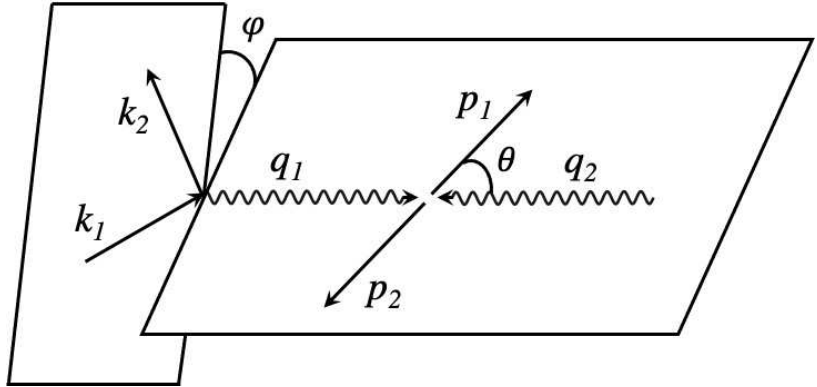}
\caption{Kinematics of the process $\gamma^*(q_1) \gamma(q_2) \to \pi(p_1) \pi(p_2)$ in the center of mass of the meson pair; the virtual photon is emitted by the electron, $q_1=k_1-k_2$. }
\label{fig:angle}
\end{figure}

To describe the process \eqref{process}, we define the lightlike vectors $n$ and $\tilde{n}$ as
\begin{align}
\tilde{n} = q_1+(1-\tau)q_2, \qquad  n=q_2,
\label{eqn:lightlike}
\end{align}
where  $\tau=s/(Q^2+s)$, $Q^2=-q_1^2$, and $s=(q_1+q_2)^2=(p_1+p_2)^2$. The polar angle of the meson ($M_i$)  momenta $\theta$ is illustrated in Fig.\,\ref{fig:angle} and  defined as ($m$ is the mass of the meson):
\begin{align}
\cos{\theta}=\frac{2q_1\cdot(p_2-p_1)}{\beta_0\,(Q^2+s)}, \qquad \beta_0=\sqrt{1-\frac{4 m^2}{s}}.
\label{eqn:polar}
\end{align}
The skewness variable 
\begin{align}
\zeta_0= \frac{(p_2-p_1)\cdot n}{(p_2+p_1)\cdot n},
\label{eqn:skewness}
\end{align}
is related to $\cos\theta$ through $\zeta_0=-\beta_0 \cos{\theta}$. In our kinematics,  only $\Delta=p_2-p_1$ has a transverse momentum,
 $\Delta=\zeta_0(\tilde{n}-\tau n) + \Delta_T$. Using the on-shell condition,  $\Delta_T^2=4m^2-(1-\zeta_0^2)s$.

The leading-twist amplitude was first presented in
Ref.~\cite{Diehl:2000uv}
with the help of a twist-2 GDA $\Phi(z,\zeta_0, s)$ for an isoscalar meson pair,
\begin{align}
\langle M(p_2) M(p_1)  | \, O_{++}(z_1 n, z_2 n) \, | 0 \rangle
= \chi\, 2 P\cdot n \int dz \,e^{2i \left[ zz_1+(1-z)z_2 \right] P\cdot n}\, \Phi(z,\zeta_0, s),
\end{align}
where $P=(p_1+ p_2)/2$, $\chi= 5e^2/18$ and $\Phi$ is the sum of GDAs  of the quark flavors $u$ and $d$, $\Phi=\Phi_u+\Phi_d$.
$O_{++}(z_1 n, z_2 n)$ is the leading-twist vector operator (a light-like Wilson line joining the points $z_1n$ and $z_2n$ is implied), 
\begin{align}
O_{++}(z_1 n, z_2 n)= \chi \left[ \bar{u}(z_1 n) \slashed{n}u(z_2 n)+ \bar{d}(z_1 n) \slashed{n}d(z_2 n) \right].
\label{eqn:opt2}
\end{align}

 The matrix element of this operator can also be expressed in terms of double distributions as 
\cite{Teryaev:2001qm}
\begin{align}
\langle M(p_2) M(p_1)  | \, O_{++}(z_1 n, z_2 n)  \,  | 0 \rangle
= \chi  \int d\beta\, d\alpha   \left[f(\beta, \alpha) \,\Delta\cdot n- g(\beta, \alpha)\, 2P\cdot n   \right] 
e^{-i  l_{z_1 z_2}\cdot n}
\label{eqn:dds}
\end{align}
with $f$ and $g$ having support on the rhombus $|\alpha|+|\beta|\leq 1$ and assumed to vanish at the boundary, and
\begin{align}
l_{z_1 z_2}=(z_2-z_1) \left[ \beta\, \frac{\Delta}{2} -(\alpha+1) P  \right] -2z_1 P.
\label{eqn:lz1z2}
\end{align}
 Then,  setting $z_1-z_2=1$, one can easily relate the GDA to double distributions 
\begin{align}
\Phi(z,\zeta_0, s)=2\int d\beta \, d\alpha\,  \delta(y+\alpha-\beta\zeta_0 ) 
\left[f(\beta, \alpha)\, \zeta_0 - g(\beta, \alpha) \right], 
\label{eqn:dd-gda}
\end{align}
where $y=2z-1$. 

In general,  GDAs can be expanded as \cite{Diehl:2000uv}
\begin{align}
\Phi(z, \cos \theta, s)=6 z(1-z) \sum_{n=1,\text{odd}}^{\infty} \sum_{l=0, \text{even}}^{n+1}
\tilde{B}_{nl}(s) C_n^{(3/2)}(2z-1) P_l(\cos \theta).
\label{eqn:gda-expression}
\end{align}
In the  asymptotic limit ($Q^2 \rightarrow \infty$), only the   $n=1$ term  survives,
\begin{align}
\Phi(z, \cos \theta, s)=18 z(1-z) (2z-1) \left[\tilde{B}_{10}(s)
+\tilde{B}_{12}(s)  P_2(\cos \theta) \right],
\label{eqn:gda-exp-asm}
\end{align}
where the first  and second terms indicate the S-wave and
D-wave production of a meson pair, respectively. 

\section{Helicity amplitudes }

The amplitude for $\gamma^* \gamma \to M M $ reads,
\begin{align}
A_{\mu \nu}=i\int d^4x\, e^{-ir\cdot x} \langle M(p_2) M(p_1)  | \, 
T \{ j_{\mu}^{\text{em}}(z_1x)  j_\nu^{\text{em}} (z_2x) \} \, | 0 \rangle , \! \!
\label{eqn:amp0}
\end{align}
where $r=z_1q_1+z_2 q_2$, and the constraint $z_1-z_2=1$ is imposed. Electromagnetic gauge invariance leads to the decomposition
 \cite{Braun:2012bg}
\begin{align}
A^{\mu \nu}=-A^{(0)}\,g_{\perp}^{\mu\nu}+A^{(1)}\, \frac{\Delta_{\alpha}g_{\perp}^{\alpha \nu}}{Q}
\left(\tilde n^\mu+(1-\tau)n^\mu\right)
+\frac{1}{2}\, A^{(2)}\, \Delta_{\alpha}\Delta_{\beta}(g_{\perp}^{\alpha \mu} g_{\perp}^{\beta \nu}- \epsilon_{\perp}^{\alpha \mu}   \epsilon_{\perp}^{\beta \nu} )+ A^{(3) \mu}\, n^{\nu}
\label{eqn:amp1}
\end{align}
 with $ g_{\perp}^{\mu \nu}$ and $\epsilon_{\perp}^{ \mu \nu}$  given by
\begin{align}
g_{\perp}^{\mu \nu} =g^{\mu\nu}-
\frac{n^{\mu}\tilde{n}^{\nu}+n^{\nu}\tilde{n}^{\mu}}{n\cdot \tilde{n}}, \qquad    
\epsilon_{\perp}^{\mu \nu} = \epsilon^{\mu \nu \alpha \beta}\,
 \frac{\tilde{n}_{\alpha} n_{\beta}}{n\cdot \tilde{n}}.
\label{eqn:gt}
\end{align}
The last term in Eq.\,(\ref{eqn:amp1}) is of no interest since it does not contribute
to any observable, and the rest of them can be expressed in terms of the GDAs 
if the factorization condition  $Q^2 \gg s, \Lambda_{\text{QCD}}^2$ is satisfied. 

To calculate the helicity amplitudes, we define the photon polarization vectors as \cite{Diehl:2000uv}
\begin{align}
&\epsilon_{0}^{\mu}=\frac{1}{Q}(|q_1^3|, 0, 0, q_1^0), \, \,
\epsilon_{\pm}^{\mu}=\frac{1}{\sqrt{2}}(0, \mp 1, -i, 0),
\label{eqn:pol-virt}
\end{align}
where the lower indices $\pm$ and $0$ indicate the helicities of the photon.
The polarization vectors $\tilde{\epsilon}$ for the real photon only have the transverse components, and they are 
related to the ones of   the virtual photon as $\tilde{\epsilon}_{\pm}=- \epsilon_{\mp}$. 
There are three independent helicity amplitudes 
$
A_{i j}=  \epsilon_{i}^{\mu} \tilde{\epsilon}_{j}^{\nu } A_{\mu \nu}.
$
We choose the independent helicity amplitudes as $A^{(0)}=A_{++}$, $A_{0+}=-A^{(1)} (\Delta \cdot \epsilon_{-})$ and $A_{-+}=-A^{(2)} (\Delta \cdot \epsilon_{-})^2$.

We shall not detail here\footnote{Details will be given in a forthcoming publication.} our calculation of the helicity amplitudes of $\gamma^* \gamma \to M M $; we adopt (and adapt to our case)   the  techniques  used for the DVCS amplitude in Ref.\,\cite{Braun:2012bg}.
Our results for the helicity amplitudes in terms of GDAs read :
\begin{align}
A^{(0)}=& \chi \left\{  (1- \frac{s}{2Q^2}) \int_0^1 dz \frac{\Phi(z, \eta, s)}{1-z} 
-\frac{s}{Q^2}\int_0^1  dz  \frac{\Phi(z,\eta, s)}{z} \ln(1-z)
\right. \nonumber \\
&-\left.  (\frac{2s}{Q^2} \eta   +\frac{\Delta_T^2}{\beta_0^2 Q^2} \frac{\partial}{\partial \eta} )
\frac{\partial}{\partial \eta}
  \int_0^1 dz \frac{ \Phi(z,\eta, s) }{z} \left[ \frac{\ln(1-z)}{2} +Li_2(1-z) -Li_2(1)  \right]
  \right \}, \nonumber \\
A^{(1)}=&  \frac{2\chi}{\beta_0 Q} \frac{\partial}{\partial  \eta }   \int_0^1 dz  \Phi(z, \eta, s) 
\frac{\ln(1-z)}{z}, \nonumber \\
A^{(2)}=&-   \frac{2 \chi}{ \beta_0^2 Q^2} \frac{\partial^2}{\partial \eta^2}       \int_0^1 dz  \Phi(z, \eta, s) 
\frac{2z-1}{z} \ln(1-z),
\label{eqn:int-gdas-a}
\end{align}
where $\eta=\cos \theta$.
The target mass correction $m^2/Q^2$ is implicit through $\Delta_T^2=4m^2-(1-\zeta_0^2)s$.

\section{Numerical estimates of kinematical higher-twist corrections to the cross section}

The process $\gamma^* \gamma \to M M $ can be measured in  $e^+ e^-$ collisions, as demonstrated  at KEKB. The cross section for $e \gamma \to e M M$ is expressed as \cite{Diehl:2000uv}
\begin{align}
\frac{d \sigma}{dQ^2 ds d(\cos \theta) d\varphi}=&
\frac{\alpha_{\text{EM}}^3 \beta_0}{16 \pi s_{e\gamma} } \frac{1}{Q^2(1-\epsilon)}   \left[ |A_{++}|^2+ |A_{-+}|^2+2\epsilon |A_{0+}|^2 -\cos \varphi  \text{Re}(A_{++}^{\ast} A_{0+} -A_{-+}^{\ast} A_{0+}) \right. \nonumber \\ 
 &\left.  + 2 \epsilon \cos (2\varphi)  \text{Re}(A_{++}^{\ast} A_{-+})   \right],
\label{eqn:epho-cro}
\end{align}
where $\varphi $ is the azimuthal angle of the meson pair as illustrated in Fig.\,(\ref{fig:angle})
and $s_{e\gamma}$ is center-of-mass energy of  
$e \gamma$. $\epsilon$ is defined as $\epsilon=\frac{1-y}{1-y+y^2/2}$, $y=\frac{Q^2+s}{s_{e \gamma}}$ and
$\alpha_{\text{EM}}=\frac{e^2}{4 \pi}  $ .

 In 2016, the Belle Collaboration released the measurements of the differential cross section for $\gamma^* + \gamma \rightarrow \pi^0+ \pi^0$\cite{Belle:2015oin}, from which the twist-2 pion GDA \cite{Kumano:2017lhr} was extracted by using the leading-twist amplitude. 
Now we adopt the  pion GDA to estimate the cross section of  $e \gamma \to e \pi  \pi $ where the integral of $\varphi$  is performed, and use this pion GDA
 to show the size of the kinematical contributions via  Eqs.\,(\ref{eqn:int-gdas-a}) and (\ref{eqn:epho-cro}). Following  the kinematics of the Belle measurements, the values of $Q^2$ are chosen as  $Q^2=9, 16, 25$ GeV$^2$, $s \in (0.25, 4)$ GeV$^2$ and we temporarily set $s_{e \gamma}=30 $ GeV$^2$ which is the typical value at Belle.

\begin{figure}[htp]
\centering
\includegraphics[width=0.9\textwidth]{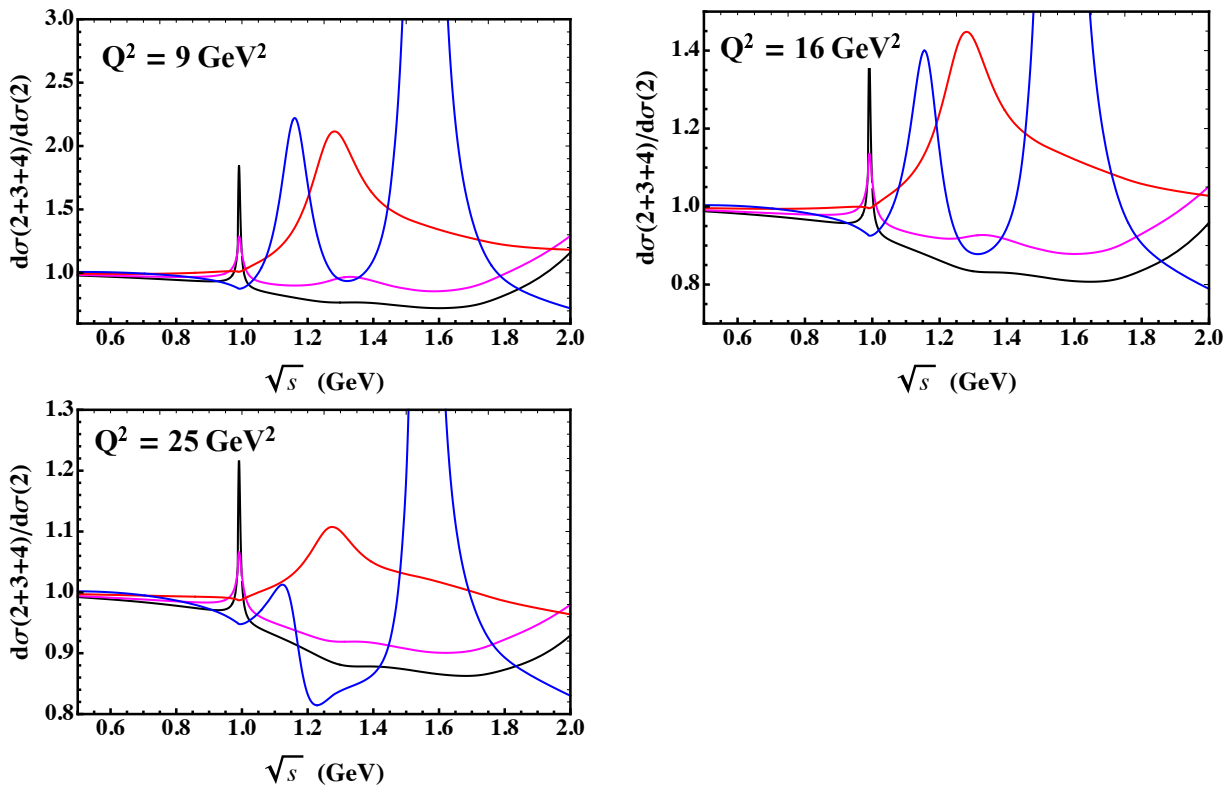}
\caption{ Ratio of $d\sigma_{t2+t3+t4}/d\sigma_{t2}$ with the pion GDA  extracted from Belle measurements.}
\label{fig:num2}
\end{figure}

In Fig.\,(\ref{fig:num2}), we  present the ratio $d\sigma_{t2+t3+t4}/d\sigma_{t2}$ where $d\sigma_{ti}$ is the kinematical twist-$i$ cross section for different values of $\cos \theta$  : black (pink, red, blue) lines denote $\cos \theta=0.2$ ($0.4, 0.6, 0.8)$. In this figure, the contributions of  the kinematical higher-twist corrections are quite clear,
and we can infer that the kinematical corrections cannot be neglected when $\sqrt{s} > 1$ GeV.
Around $\sqrt{s} \sim 1.5$ GeV, the kinematical corrections are dominant in the cross section with $\cos \theta=0.8$, but this is mainly due to the fact that the twist-2 cross section is  tiny with the extracted GDA from Belle measurements; however, this GDA may not be accurate enough since the uncertainties of Belle measurements are large in these kinematics.
As $Q^2$ increases, the role of the kinematical contributions becomes less important,
 consistently with the fact that the higher-twist kinematical contributions  are suppressed by $1/Q$ or $1/Q^2$.

To sum-up this phenomenological study, let us stress that the uncertainties  of Belle measurements \cite{Belle:2015oin} are quite large, and the statistical errors are dominant.
However, this situation will be improved substantially soon, since Belle II collaboration just started taking data at the SuperKEKB with a much higher luminosity. Precise measurements of $\gamma^* + \gamma \rightarrow M_1+ M_2$ are expected in the near future, and an accurate description of the amplitudes will require the inclusion of kinematical contributions up to twist 4. This will be of utmost importance to address the question of the form factors of the pion energy-momentum tensor \cite{Kumano:2017lhr}
 and of  the impact-parameter picture representation of GDAs \cite{Pire:2002ut}.

\end{document}